\documentstyle[12pt,prl,aps]{revtex}

\draft
\begin{document}
\title{
A Rapid Dynamical Monte Carlo Algorithm for Glassy Systems}
\author{Werner Krauth$^*$ and Olivier Pluchery$^{**}$}
\address{
$^{*}$ CNRS-Laboratoire de Physique Statistique de l'ENS\\
24, rue Lhomond; F-75231 Paris Cedex 05; France\\
e-mail: krauth@physique.ens.fr\\
$^{**}$ Ecole Normale Sup\'{e}rieure de Cachan\\
61, av. du Pr\'{e}sident Wilson; F-94235 Cachan Cedex; France\\
}
\date{July 1994}
\maketitle
\begin{abstract}
In this paper we present a dynamical Monte Carlo algorithm which is applicable 
to systems satisfying a clustering condition: 
during the dynamical evolution the system is mostly trapped in deep 
local minima (as happens in glasses, pinning problems etc.).  We compare the
algorithm to the usual Monte Carlo algorithm, using as an example the 
Bernasconi model. In this model, a straightforward implementation of the
algorithm gives an improvement of several orders of magnitude
in computational speed 
with respect to a recent, already very efficient, implementation of the 
algorithm of Bortz, Kalos and Lebowitz.
\end{abstract}
\newpage
Direct dynamical simulations have traditionally played an important 
role in statistical physics. Investigations have covered a large range 
of problems, from dynamical phase transitions (e.g. the spinodal 
decomposition and the glass transition) to detailed, realistic simulations
of polymers and long proteins in the sub-nanosecond regime.

The macroscopic evolution of the system is always much slower than
its microscopic Monte Carlo dynamics. This simply means that the 
problem of dynamical simulations is, almost by definition, a very difficult 
enterprise.

In dynamical Monte Carlo simulations one considers Markov sequences of 
configurations
\begin{equation}
\sigma_1(\tau_1) \rightarrow \sigma_2(\tau_2) 
 \ldots \rightarrow \sigma_i(\tau_i) 
 \rightarrow \sigma_{i+1}(\tau_{i+1})\ldots
etc
\label{MCsequence}
\end{equation}
where the symbols mean that the system will remain 
in the configuration $\sigma_i$ during the time $\tau_i$. For concreteness
we imagine the configuration to stem from a spin model 
$\sigma = (S_1, S_2,\ldots,S_N)$ with $S_i=\pm1$. In the single-spin
flip Metropolis dynamics, the system can move from configurations
$\sigma_i$ to any configuration which differs from it by the flip 
of a single spin $m$, which we denote by $\sigma^{[m]}$.
The probability of making a move is then given by
\begin{equation}
p(\sigma \rightarrow \sigma^{[m]},\Delta\tau) = \frac{\Delta\tau}{N}
\left\{ \begin{array}{l}
1 \;\;\;\;\;\;\;\;\;\;\;\;\;\;\; (if\;\;\; E(\sigma^{[m]}) < E(\sigma)) \\
\exp[-\beta( E(\sigma^{[m]}) - E(\sigma))]\;\;\ (otherwise)
\end{array} \right. m=1,\ldots, N
\label{transition}
\end{equation}
where $E(\sigma)$ is the energy of the configuration $\sigma$.
Usually, the sequence in eq. (\ref{MCsequence})
is directly simulated (with a {\it finite} timestep $\Delta\tau$) 
by means of
a rejection method. In that case, the time intervals $\tau_i$ in
eq. (\ref{MCsequence}) are multiples of $\Delta\tau$:  
$\tau_i =M\Delta\tau$.
At each time step a flip of a randomly chosen
spin $m$ is attempted, and a uniform random number is drawn. Then
\begin{equation}
      \sigma(\tau + \Delta\tau)=\{ \begin{array}{l}
\sigma^{[m]} \;\;\;\; if\;\;\;\;\;\;\; p(\sigma 
\rightarrow \sigma^{[m]}) >ran\;\;\;0<ran<1\\
\sigma\;\;\ (otherwise) \end{array} 
\label{rejection}
\end{equation}

At low temperatures, the MC dynamics often encounters the problem of small 
acceptance
probabilities: as soon as the temperature is such that the available 
phase space is dominated by low-lying minima, the system will take 
a long time to move from a low-lying state to an excited one. 
The problem of small acceptance probabilities is certainly a nuisance, 
but not a true impediment to the simulation. Almost 20 years ago, Bortz,
Kalos and Lebowitz \cite{bortz} (BKL) showed how to circumvent it: 
in a nutshell, their algorithm consists
in giving up the rejection method in favor of a direct calculation of
waiting times, which can be calculated from the knowledge of all the
transitions. In the above example, the time $\tau_i$ the system will 
stay in a configuration $ \sigma_i$ can be directly sampled from its 
probability distribution $p(\tau_i)=\lambda \exp(-\lambda \tau_i)$ with
$\lambda = \sum_{m=1,N} p(\sigma_i \rightarrow \sigma_i^{[m]})$.
The BKL algorithm 
has been used repeatedly ({\it e.g.} in \cite{sethna} \cite{kinzel}) 
over the last years.  A recent, very efficient
implementation \cite{KM2} has given improvements in computational speed 
of the order of $10^3$ to $10^5$ for glassy systems at low temperatures.

If the problem of small acceptance probabilities has thus been effectively
solved by the BKL algorithm, it is usually accompanied by another
one, which we may call ``futile'' dynamics: if the dynamical evolution 
is dominated by isolated deep local minima, the system will get
trapped even though we force it to make transitions out of them: very shortly 
after having escaped a minimal configuration (by means of the BKL 
algorithm),
the system simply falls back into it. This means that, at low temperature,
the dynamics consists mostly of short cycles.

The purpose of the present paper is to show that this problem, also, can
be eliminated, and the power of the BKL algorithm increased
by at least a few orders of magnitude. To give an illustration of the 
problem we are concerned with and to introduce the model 
which will serve as an
example for our presentation, we show in fig. 1 the result of a very long 
simulation of a single run of the Bernasconi model \cite{bernasconi}, defined
by the hamiltonian:
\begin{equation}
{\cal H} = {1 \over N} \sum_{k=1}^{N-1} \left [\sum_{i=1}^{N-k} S_i
S_{i+k}\right]^2  
\label{bernasconi}
\end{equation}
The model defined in eq. (\ref{bernasconi}) has acquired some fame 
during the last years 
as an example of a ``tough'' optimization problem. An inkling of why this is
so 
can immediately be gotten from fig. 1, in which we show the 
temporal evolution 
of a system of size $N=400$ at a temperature $T=0.05$. The simulation 
has gone over a time $t=3.9\times 10^8$, 
({\it i.e.} has performed the equivalent
of $3.9\times 10^8$ Monte Carlo steps {\it per spin}). 
The total number of accepted moves
is around $N_{acc}=2.1\times 10^9$,  but the total number of different
configurations visited, $N_{diff} = 1290$, is very small.
In an effort to be completely explicit,
we restate that the small acceptance probability ($N_{acc}/(N*t)\ll 1$) is
related but not identical to the ``futility'' expressed by 
$N_{diff}/N_{acc}\ll 1$.  On the second curve in fig. 1
we display the total number of different configurations visited as a function 
of time. One can clearly see that this number grows very slowly
whenever the system is within a given plateau of the energy.
Many of the configurations are thus visited
a large number of times. It should be evident that the total
time the system spends in any of these configurations
can be evaluated in a much more efficient
way than by visiting them repeatedly. It can in fact be deduced
from the Boltzmann distribution. We have extensively tested this  
fundamental hypothesis of our method in simulations using the BKL algorithm.

Consider now the schematic view of the algorithm presented in fig. 2.
This algorithm completely avoids the computational slowdown associated
with repeated visits to the same configuration. 
In fact, it visits one {\it new} configuration per iteration.
At any given time, there is an active set of configurations 
which make up the ``cluster'' ${\cal C}$. For any configuration $i$ in 
${\cal C}$, we
have stored, besides the energy E(i),  the total probability to move from
site $i$
to a configuration  outside the cluster 
\begin{equation}
p_{ext}(i)=\sum_{j \not \epsilon {\cal C}}p(i\rightarrow j)
\end{equation}
Now we suppose that the configurations in ${\cal C}$
are in quasi-equilibrium, {\it i.e.} we assume that 
the system visits configuration $j$ with the probability given 
by its normalized Boltzmann weight
\begin{equation}
p_{eq}(i)=  \frac{\exp(-\beta E(i))}{\sum_{k \epsilon{\cal C} 
}\exp(-\beta E(k))}
\end{equation}
The (normalized) probability to move from any configuration in {\cal C}
to a configuration outside {\cal C} is thus given by
$\sum_{i \epsilon {\cal C}}p_{eq}(i)p_{ext}(i)$. In direct analogy
to the BKL algorithm, we then sample the time $\tau$ the system will stay in 
${\cal C}$  from its probability distribution
\begin{equation} 
p(\tau) = \lambda \exp(-\lambda \tau)\;\;\; with \;\;\;\lambda =  
\sum_{i\epsilon{\cal C}}p_{eq}(i) p_{ext}(i)
\label{biglambda}
\end{equation} 
As before, $\lambda$ is the parameter of the exponential distribution
of waiting times in the cluster. It should be evident that $\lambda$
can very quickly become a small quantity, which needs to be calculated
with high-precision arithmetic. In the calculation presented later on 
(fig. 4),
$\lambda^{-1}$ routinely  takes on values of $\lambda^{-1} \sim 3 
  \times 10^9$, 
which simply means that a {\it single } iteration of the algorithm
(at very large times of the simulation) corresponds to $3 \times 10^9$ 
Monte Carlo steps per spin! 
After this time, sampled from eq. (\ref{biglambda}), 
the system moves outside ${\cal C}$. 
Now, in analogy with the BKL algorithm, we are able to sample in its turn
the configuration of ${\cal C}$ (denoted by {\bf A} in fig. 2) from 
which this move will be made (this probability is simply proportional 
to $p_{eq}(j) p_{ext}(j)$). Finally, having identified the configuration
$\sigma_i$ ($\equiv$ {\bf A}), we can sample the configuration outside 
${\cal C}$, according to eq. (\ref{transition}), after having excluded
all the moves which stay within ${\cal C}$. This then produces the
configuration denoted by {\bf B} in fig. 2.

In our current implementation of the algorithm, we then calculate
the probability $p_{cluster}$ to return to the cluster from {\bf B}, 
or to move 
directly to  configurations outside the cluster, $p_{ext}$. 
In both cases, the amount of time spent on {\bf B} is sampled 
from $p_{self}= 1-p_{ext} - p_{cluster}$. 
We then add {\bf B} to the 
cluster and update all the probabilities $p_{ext}(i)$ for 
all members of ${\cal C}$, whose size, in the meantime, has been incremented.

A little thought shows that the algorithm just sketched can be implemented
recursively, and that a single iteration consists in two passes through
the cluster. Furthermore, besides the configurations (which we keep
in external memory), we only need the tables $E(i)$ and $p_{ext}(i)$,
{\it i.e.} there is no need for tables which grow faster than 
the cluster size. In the example treated, we can presently, 
for $N=400$, allow up to $10^4$ members in ${\cal C}$.

To show that the algorithm works, we have tested it against the BKL algorithm.
In fact, for both algorithms, we have started 300 samples on the same 
configuration (marked ``{\bf X}'' in fig. 1), and have determined the  
histogram
of trapping times (times after which the system accesses
a new plateau with lower energy). This histogram is shown in fig. 3. 
As can be clearly seen, the distributions
of waiting times are identical, which indicates that
the algorithm is correct for all intents and purposes. Furthermore, it
should be evident that, once an old plateau has been left, and a new
configuration of minimal energy found, the original members of 
${\cal C}$ drop out of the picture. At this point, a new cluster can 
be started. 

In fig. 4  we show the result of a simulation for the same system as
in fig. 1, but this time with the new algorithm. Using the same amount
of computer time, we reach physical times which are about $10000$ times
larger. The inset in fig. 4 shows the energy of the new configurations
which enter the cluster. This energy drops as a new plateau is found,
and then increases (approximately logarithmically in time), as more
and more remote configurations are encountered  during
the thermal exploration process. The main plot in fig. 4
shows the mean energy of the cluster. Note that many more quantities
are accessible, such as correlation functions, or more detailed 
properties of the dynamical process. 

Finally we would like to discuss whether the algorithm, as presented, and
as sketched in fig. 2 can be made rigorous. In the present state, we 
increase the size of the cluster by one at any iteration. 
Clearly, a more rigorous approach would 
be to add point ``{\bf B}'' to the cluster only if $p_{ext} \ll p_{int}$,
say if  $p_{ext}/p_{int}<\gamma$. 
The algorithm is equivalent to the BKL algorithm for $\gamma =0$, but
certainly retains some asymptotic validity as 
$\gamma \stackrel{\sim}{>}0$. The issue at hand is that the
configurations in the cluster should satisfy a quasi-equilibrium
condition, which basically means $p_{ext}\ll p_{int}$, and which may 
be implemented with various degrees of  rigor.

A second question of course concerns the criterium for abandoning
the cluster and growing a new one. In our present case we have simply
adopted 
an energy criterium: each time a new lowest energy was found, the old
cluster was abandoned, and a new one grown. The precise criterium 
will depend on the particular problem. Whenever
geometry plays an important role, the cluster may have to be scrapped
as soon as the geometrical distances between members of the cluster 
become too large.

In conclusion, we have presented in this paper an algorithm for
the Monte Carlo dynamics of glassy systems, which is
able to produce enormous gains in efficiency if 
configurations are visited over and over again.
In a natural way, the algorithm forms a cluster of a
certain number of configurations, which are close in energy. These
configurations are then assumed to be in thermal equilibrium, an 
assumption which was very well satisfied in the example we treated.

Finally, we indicate that a FORTRAN source code for the 
algorithms used in this work may be obtained by  email.

\acknowledgments
We acknowledge helpful discussions with M. M\'{e}zard.

\noindent
\newpage
{\bf Figure Captions}

\begin{enumerate}
\item
\label{evolution}
Energy {\it vs} $\log t$ for the Bernasconi model at $T=0.05$ for a single 
realization. The calculation was performed using the BKL algorithm 
in a very efficient implementation \cite{KM2}.
The point marked ``{\bf X}'' serves as a starting point for subsequent 
simulations ({\it cf} fig. 3). The second curve gives the
total number of different configurations visited (scale to
the right). When on a ``plateau'', the number of configurations 
increases roughly linearly in $\log t$.

\item
\label{schema} 
Schematic view of the Cluster Monte Carlo algorithm presented in this
paper.

\item
\label{comparison}
Histograms of trapping times for the two algorithms.
For this figure, both the BKL algorithm and the cluster algorithm presented
in this paper were started
at the point marked ``{\bf X}'' in fig. 1. 300 runs of both algorithms 
were done, and the time after which a new plateau of lower energy was
found was typically between $t=10^4$ and $t=10^6$.

\item
\label{turbo}
Mean energy of the cluster {\it vs} $\log t$ for the Bernasconi model 
at $T=0.05$ (single run of the cluster algorithm).
The  inset displays the energy of new configurations which enter the cluster. 

\end{enumerate}
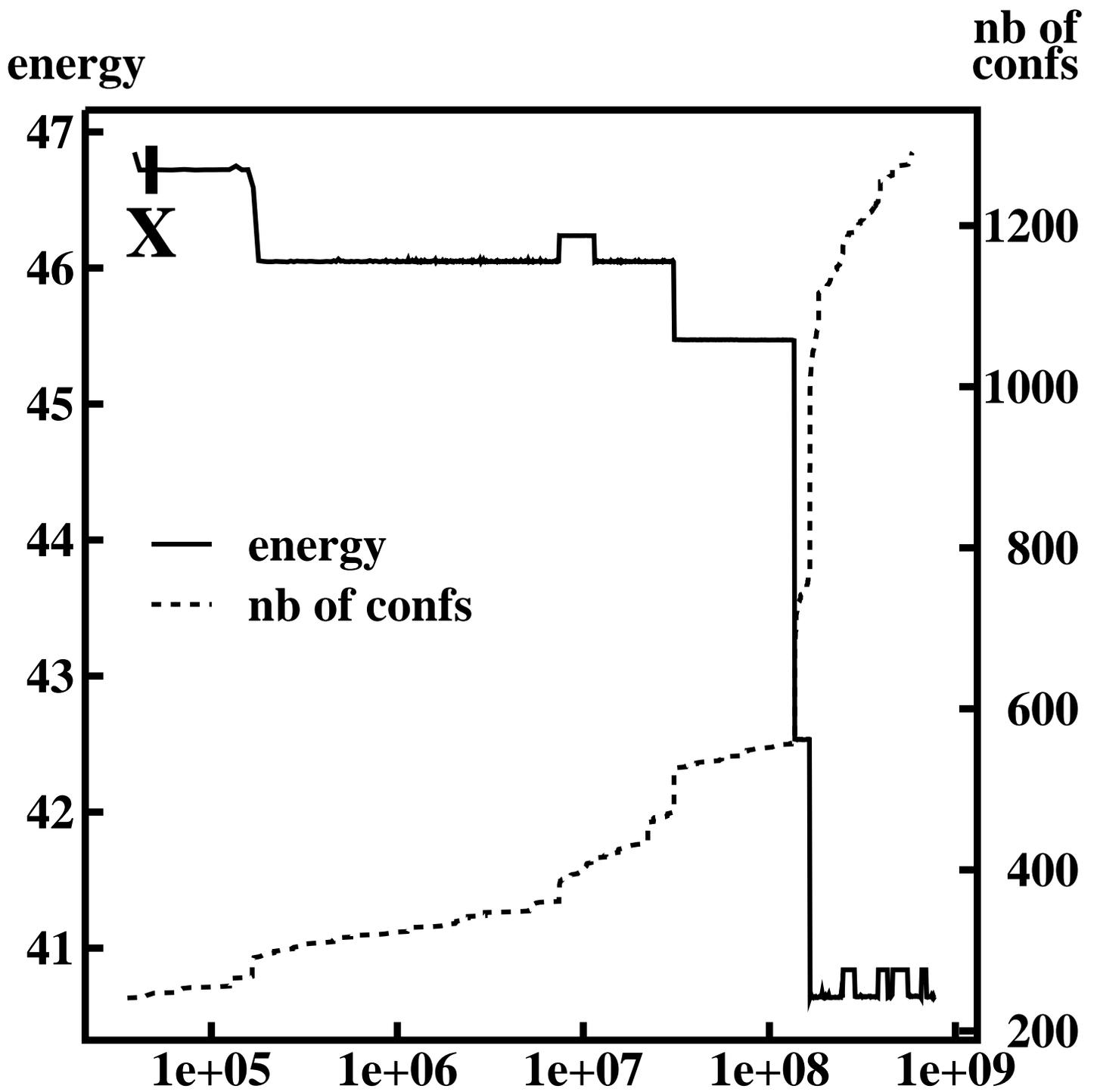
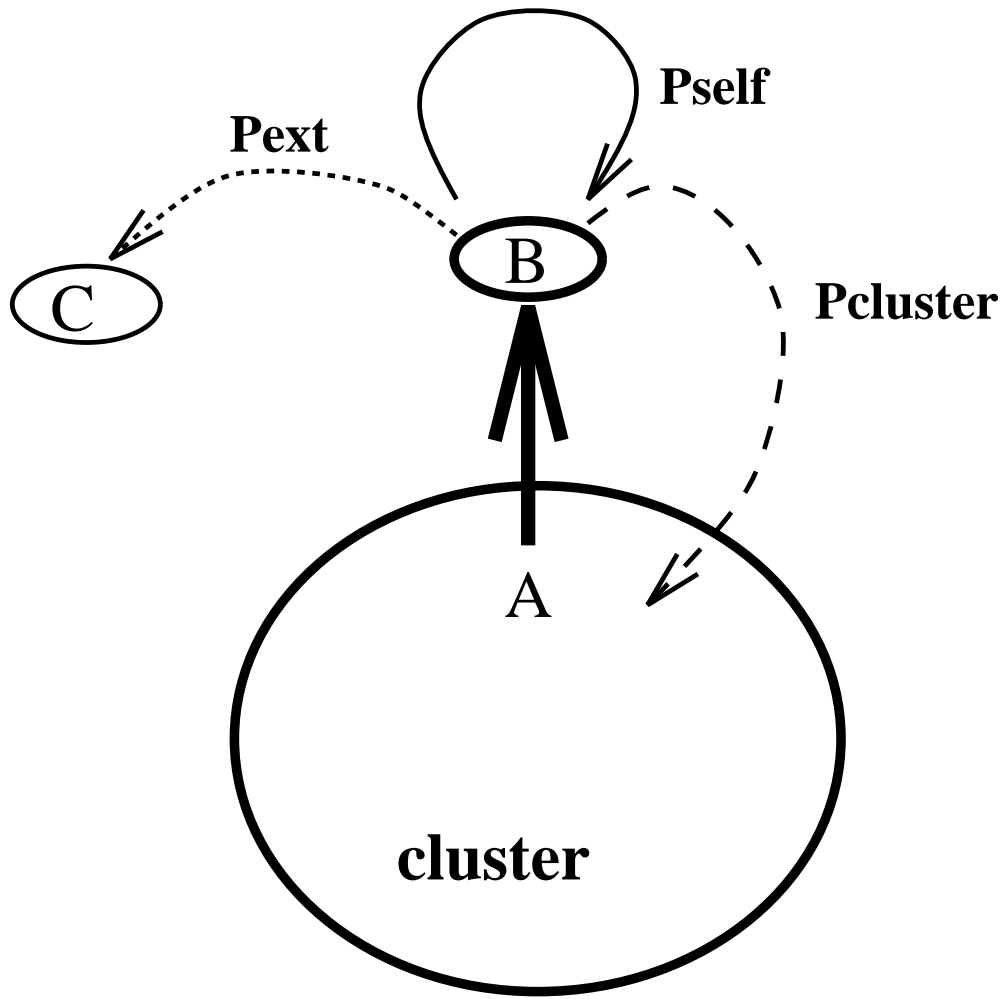
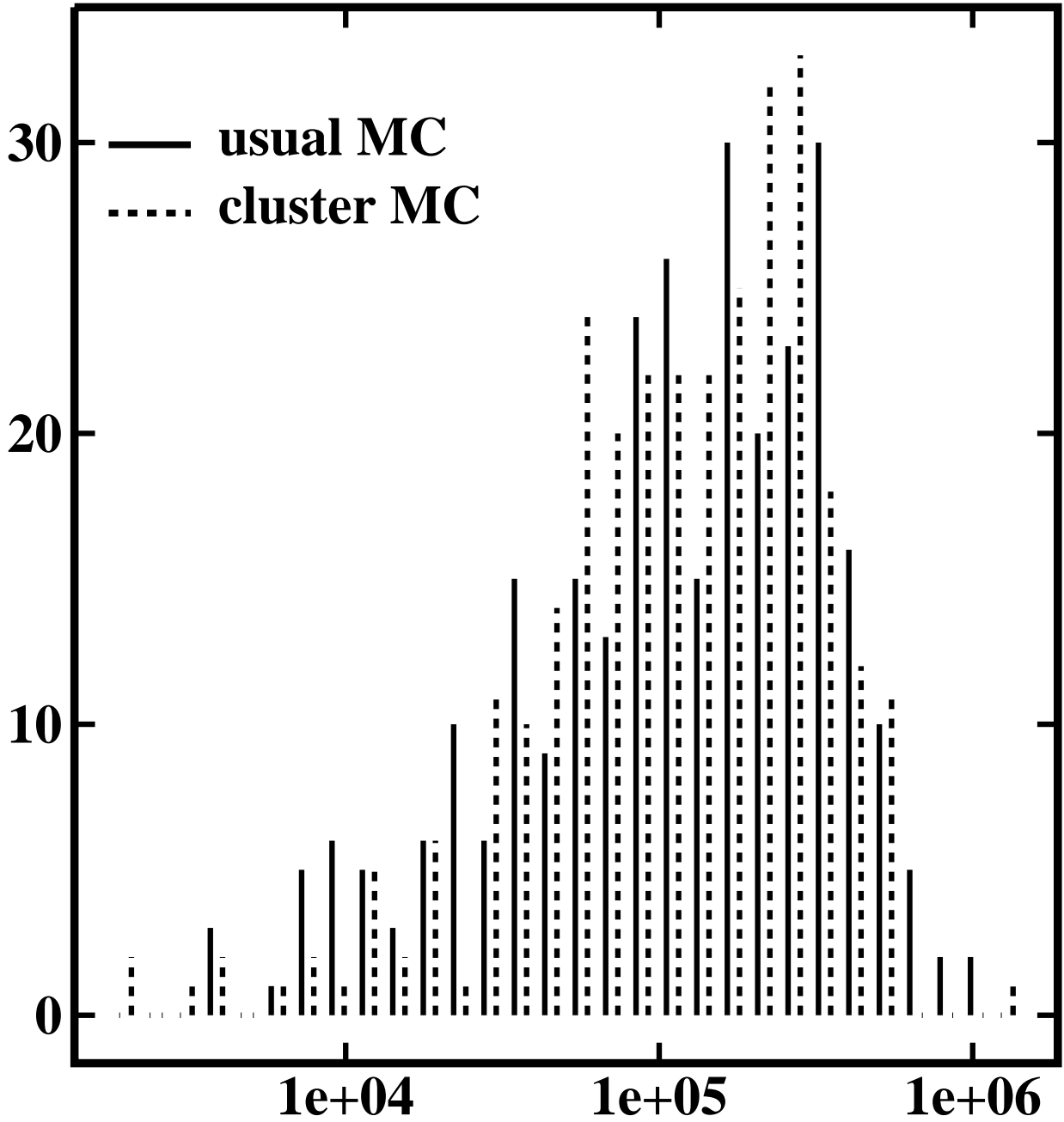
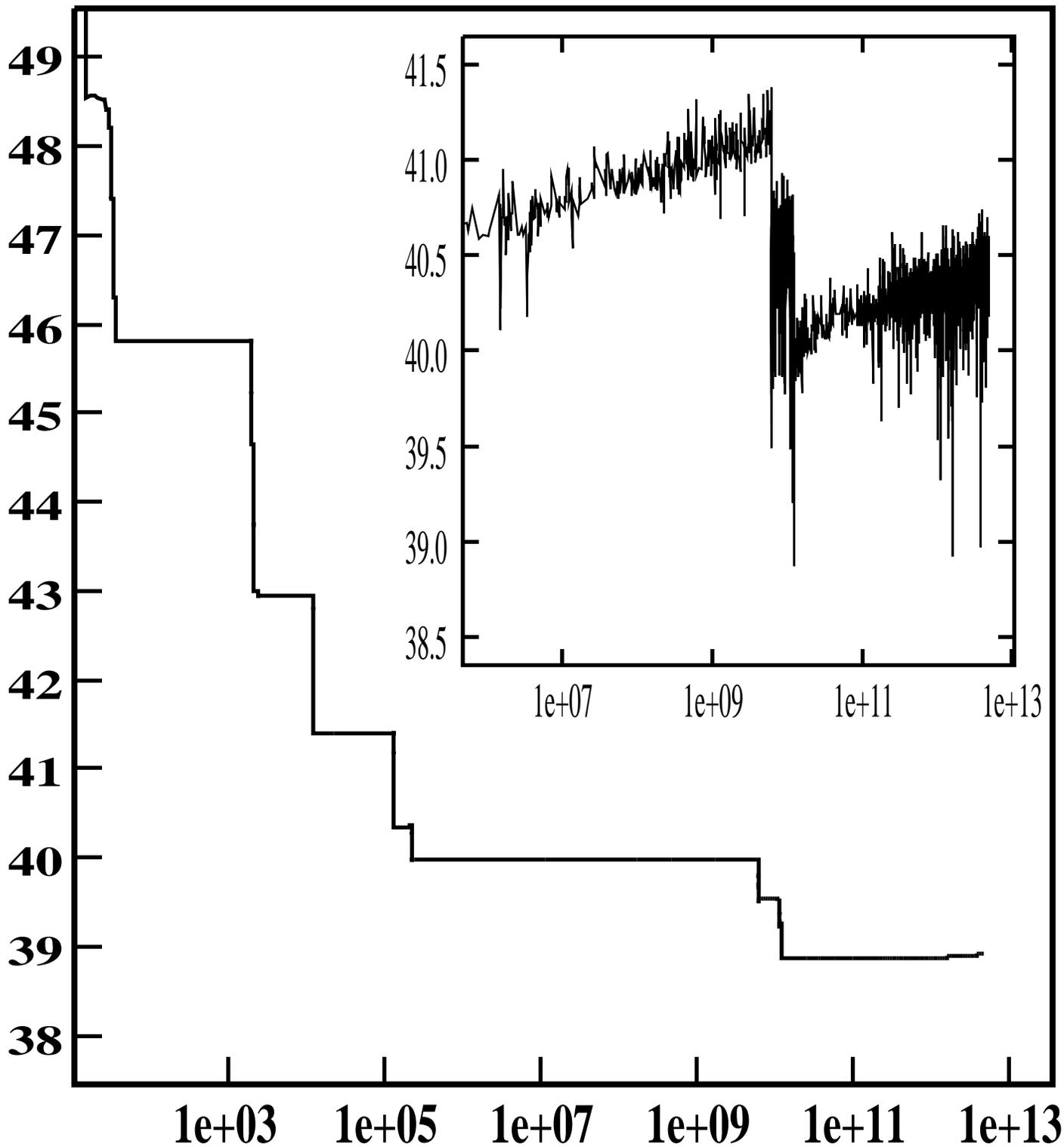
\end{document}